\documentclass[aps,pra,showpacs,amsmath,amssymb,superscriptaddress,reprint,10pt,longbibliography]{revtex4-1}
\usepackage{graphicx,mathrsfs,enumerate}
\usepackage[utf8]{inputenc}
\usepackage{mathtools}
\usepackage{amsfonts,amstext}
\usepackage{amsmath,amsthm,amssymb,mathbbol,tensor}
\usepackage{graphicx,color,mathrsfs,enumerate}
\usepackage[hidelinks]{hyperref}\hypersetup{pdfpagemode=UseNone,pdfstartview=}
\usepackage[left=2cm,right=2cm]{geometry}
\usepackage{xcolor}
\usepackage{url}
\usepackage{times}
\usepackage{bbm}
\usepackage[english]{babel}

\usepackage{chngcntr}

\DeclareMathOperator{\Tr}{Tr}

\newtheorem{thm}{Theorem}

\newtheorem{lem}[thm]{Lemma}
\newtheorem{crit}{Criterion}

\begin{document}
\title{Necessary and sufficient conditions for multipartite Bell violations  with only one trusted device}
\author{M. M. Taddei}\email{marciotaddei@if.ufrj.br}
\affiliation{Instituto de F\'isica, Universidade Federal do Rio de Janeiro, Caixa Postal 68528, Rio de Janeiro, RJ 21941-972, Brazil}
\author{R. V. Nery}
\affiliation{Instituto de F\'isica, Universidade Federal do Rio de Janeiro, Caixa Postal 68528, Rio de Janeiro, RJ 21941-972, Brazil}
\author{L. Aolita}
\affiliation{Instituto de F\'isica, Universidade Federal do Rio de Janeiro, Caixa Postal 68528, Rio de Janeiro, RJ 21941-972, Brazil} 
\date{\today}
\begin{abstract}
We study multipartite Bell nonlocality in a framework native of multipartite Einstein-Podolsky-Rosen (EPR) steering scenarios with a single trusted measurement device. We derive a closed-form necessary and sufficient criterion for systems composed of a qubit and $N-1$ untrusted black-box measurement devices to violate -- under general dichotomic measurements on the qubit -- a generic Bell inequality from a broad family of linear inequalities with arbitrarily many outputs for the $N-1$ untrusted devices and inputs for all $N$ parties. The optimal quantum measurements for maximal violation are also obtained. For two users, and two inputs and two outputs per user, our criterion becomes necessary and sufficient for Bell nonlocality.
Furthermore, in that setting, its form generalizes recently obtained steering inequalities, which allows us to provide useful feedback from nonlocality to the detection of steering. 
Our findings constitute a practical tool for the study of the interplay between EPR steering and Bell nonlocality, with potential applications in multipartite information processing.
\end{abstract}
\maketitle

{\it Introduction}.--- Composite quantum systems can
display exotic forms of non-classical correlations, a phenomenon known under the generic name of \emph{quantum nonlocality}. Quantum nonlocal correlations can appear in three main variants. The first one is \emph{entanglement}, which refers to inseparability of quantum states (described by density matrices) \cite{Horodecki09}. The second one is {\it Bell nonlocality} \cite{Brunner13}, the impossibility of explaining measurement statistics (described by joint probability distributions) with local hidden-variable (LHV) models \cite{Bell}. The third variant is called {\it Einstein-Podolsky-Rosen (EPR) Steering} \cite{Reid09}, after the famous EPR paper \cite{EPR}. This is an effect by which ensembles of quantum states are remotely prepared by local measurements at distant labs \cite{Reid09}. 
The observable data in steering experiments thus consist of the measurement statistics at one lab and quantum states at another. These data are compactly described by a joint mathematical object called \emph{assemblage}, composed of a conditional probability distribution and a set of density matrices. Hence, EPR steering constitutes an intermediate notion between entanglement and Bell nonlocality \cite{Wiseman07}.

Apart from their fundamental importance, nonlocal correlations are of practical relevance: they are resources for physical tasks such as quantum key-distribution (QKD) \cite{Deviceindependent_QKD,Branciard12} and quantum random-number generation \cite{Deviceindependent_QRNG}. Entanglement is a resource \cite{Horodecki09} in the device-dependent (DD) scenario, defined by well-characterized, trusted quantum measurements. Bell nonlocality is useful for device-independent (DI) protocols \cite{Deviceindependent_QKD, Deviceindependent_QRNG}, i.e., where the experimenters possess untrusted black-box measurement apparatuses. In turn, EPR steering has been identified \cite{Branciard12} as a resource \cite{Gallego15} for one-sided DI situations, where one of the parties has an untrusted black-box device while the other possesses a trusted quantum platform. Fully (both-sided) DI protocols relax the need for device characterization totally, but at the expense of being very demanding experimentally \cite{LoopholefreeBell}. 
One-sided DI implementations offer a middle-path alternative, relaxing device characterization only on one side but having, in return, less stringent experimental requirements \cite{Wittmann12} for security \cite{Branciard12} than in fully DI ones. This is relevant to any asymmetric situation involving users with different levels of quantum control.

These developments motivated a great amount of work on the interplay between the different forms of quantum nonlocality. The first problem tackled was that of entangled versus Bell nonlocal states (those capable of exhibiting Bell nonlocality). All pure entangled states were proven to be Bell nonlocal \cite{Capasso+73, Gisin91, Home-Selleri91,PR92}, but mixed Bell local entangled states were found \cite{Werner89}. Later, necessary and sufficient conditions for arbitrary 2-qubit states to be Bell nonlocal were derived \cite{Horodecki95}. A long list of works then followed these pioneering results (see, for instance, Sec. III.A of \cite{Brunner13} and Refs. therein). The second problem was that of entangled  versus steerable states (those  capable of exhibiting EPR steering). Unsteerable entangled states, as well as Bell local steerable ones, were found \cite{Wiseman07, Saunders10}. This led to an impressive amount of work: the sets of entangled, steerable and Bell nonlocal states were proven inequivalent under general measurements \cite{MTQuintino}; necessary criteria for a two-qubit state to be steerable were found \cite{Bowles}; and constructive methods to test for unsteerability of a state were developed \cite{HirschCavalcantiConstruct}. Finally, in addition to the many known steering inequalities \cite{Cavalcanti09,Wittmann12, Skrzypczyk14,ECavalcanti14,Piani15,Kogias15, ECavalcanti16}, a necessary and sufficient criterion for EPR steering has been recently obtained for minimal-dimension assemblages \cite{Uola15}.

Here we consider a third problem: steerable versus Bell nonlocal multipartite assemblages. We derive a closed-form necessary and sufficient criterion for an $N$-partite assemblage with a single trusted device, in possession of a user called Alice, to violate, under general measurements [positive operator-valued measure (POVM)] by her, a Bell inequality. The optimal measurements for the maximal violation are also given.
Our theorems assume that Alice's measurements are dichotomic and that her system is a qubit, but are otherwise general in the number of outcomes for the untrusted devices and of settings for all parties. Furthermore, we make only minimal assumptions on the Bell inequalities treated, namely, that they are linear and that their violations do not increase under probabilistic local mixings of Alice's outputs.
Thus, many of the most popular bipartite \cite{CHSH,Pearle70,BraunsteinCaves90,I3322} and multipartite \cite{MABK,Svetlichny,generalSvetlichny,Aolita12,Bancal,Pal} 
Bell inequalities are within the range of applicability of our criterion.
In addition, for $N=2$ users with 2 inputs and 2 outputs per user, 
our criterion unambiguously characterizes all Bell nonlocal assemblages.
Interestingly, in that setting, our criterion generalizes, in form, recently obtained steering inequalities \cite{ECavalcanti14,ECavalcanti16}. By virtue of this, we provide insight into the detection of EPR steering within the framework of Bell nonlocality and explain formal links between the two problems. 
Finally, we suggest potential connections of our findings with information-theoretic protocols with asymmetric levels of quantum control among the users involved.

{\it Preliminaries}.--- We consider $N$ space-like separated parties, Alice, in possession of a trusted measurement device, and $N-1$ users, $B_1$, $\hdots$, $B_{N-1}$, in possession of untrusted devices \cite{Cavalcanti15}. This is the $(N-1)$-sided DI scenario. Each $i$-th untrusted device, for $i=1,\hdots, N-1$, is treated as a black box with unknown internal functioning, which, given an input $y_i\in[m]$, outputs $b_i\in[o]$, where $m,\, o\in\mathbb{N}$ and the notation $[n]\coloneqq\{0,\ldots,n-1\}$, for any $n \in\mathbb{N}$, is introduced.  In addition, we will also use the short-hand notation $[\boldsymbol{n}]\coloneqq[n]^{N-1}$. 
Alice's subsystem, in turn, is a qubit, on which she can perform any quantum measurement of her choice.
The joint system state is specified by an $(N-1)$-partite conditional probability distribution $P(\boldsymbol{b}|\boldsymbol{y})$ of the output string $\boldsymbol{b}\coloneqq b_1,\hdots,b_{N-1}$ given the  input string $\boldsymbol{y}\coloneqq y_1,\hdots,y_{N-1}$, associated to a  (normalized) conditional single-partite quantum state $\varrho_{\boldsymbol{b}|\boldsymbol{y}}
$ on  Alice's subsystem's Hilbert space $\mathcal{H}$. These can be conveniently encapsulated in the
 {\it assemblage}
\begin{equation}
\label{eq:ensembledef}
\boldsymbol{\Xi}\coloneqq\{\sigma_{\boldsymbol{b}|\boldsymbol{y}}\}_{\boldsymbol{b}\in[\boldsymbol{o}],\, \boldsymbol{y}\in[\boldsymbol{m}]},
\end{equation} 
of (subnormalized) conditional quantum states $\sigma_{\boldsymbol{b}|\boldsymbol{y}}
$ on $\mathcal{H}$, with $\sigma_{\boldsymbol{b}|\boldsymbol{y}}\coloneqq P(\boldsymbol{b}|\boldsymbol{y}) \ \varrho_{\boldsymbol{b}|\boldsymbol{y}}$. In other words, $\boldsymbol{\Xi}$ provides a concise description of all the observable information in $(N-1)$-sided DI experiments. 

On the other hand, in the fully DI scenario of all $N$ users possessing black-box devices, the joint system behavior is described by an $N$-partite conditional distribution 
\begin{equation}
\label{def:Prob_dist}
\boldsymbol{P}:=\{P(a,\boldsymbol{b}|x,\boldsymbol{y})\}_{a\in[o],\, \boldsymbol{b}\in[\boldsymbol{o}], \, x\in[m],\, \boldsymbol{y}\in[\boldsymbol{m}]},
\end{equation}
where $P(a,\boldsymbol{b}|x,\boldsymbol{y})$ is the probability of output values $a$ and $\boldsymbol{b}$ conditioned on input values $x$ and $\boldsymbol{y}$. For ease of notation, we assume throughout that the numbers of inputs and outputs, $m$ and $o$, respectively, are the same for all $N$ users, but all our results are also valid otherwise.

Bell inequalities offer a practical tool to test for Bell nonlocality in a given distribution \cite{Brunner13}. Every linear Bell inequality is represented by a pair $\{\boldsymbol{\beta},\beta_{\rm L}\}$, with $\boldsymbol{\beta}\coloneqq\{\beta_{a,\boldsymbol{b},x,\boldsymbol{y}}\in\mathbb{R}\}_{a\in[o],\, \boldsymbol{b}\in[\boldsymbol{o}], \, x\in[m],\, \boldsymbol{y}\in[\boldsymbol{m}]}$
 and $\beta_{\rm L}\in\mathbb{R}$, such that
\begin{equation}
\label{eq:Bell_Ineq}
\boldsymbol{\beta}\cdot\boldsymbol{P}\coloneqq
\sum_{\substack{  a\in[o],\, \boldsymbol{b}\in[\boldsymbol{o}]\\ x\in[m],\, \boldsymbol{y}\in[\boldsymbol{m}] }}
 \beta_{a,\boldsymbol{b},x,\boldsymbol{y}} \,  P(a,\boldsymbol{b}|x,\boldsymbol{y}) \leq \beta_{\rm L}
\end{equation} 
for all Bell local $\boldsymbol{P}$. Furthermore, in the multipartite scenario, Bell inequalities can also be tailored so as to test for different forms of multipartite Bell non locality \cite{Brunner13}. 

Our criterion below holds for all linear Bell inequalities whose violations do not increase under local probabilistic mixings of Alice's outputs, to which we refer, for short, as \emph{well-behaved Bell inequalities}. More precisely,
local mixings map $\boldsymbol{P}$ into a distribution $\boldsymbol{P}_{\rm lm}$ with elements
\begin{equation}
P_{\rm lm}(a,\boldsymbol{b}|x,\boldsymbol{y})= \sum_{a'\in[o]} q(a|a',x) \ P(a',\boldsymbol{b}|x,\boldsymbol{y}) \ ,
\label{eq:defLocMix}
\end{equation}
where $q(a|a',x)\geq0$, with $\sum_{a'\in[o]} q(a|a',x)=1$, characterizes the mixing probability for each input $x$ and output $a'$. Hence, $\{\boldsymbol{\beta},\beta_{\rm L}\}$ is well behaved if $\boldsymbol{\beta}\cdot\boldsymbol{P}_{\rm lm}\leq\boldsymbol{\beta}\cdot\boldsymbol{P}$ for all $\boldsymbol{P}$ for which $\boldsymbol{\beta}\cdot\boldsymbol{P}>\beta_{\rm L}$.
Local mixings can map local distributions only into local distributions. So, that a Bell violation does not increase under such mixings is a basic reasonable property typically satisfied by known inequalities (including all tight  ones and, more generally, all those for which a constant local weight \cite{EPR92} implies a constant violation). Examples not satisfying this property can be found among reducible inequalities that have superfluous terms \cite{GaruccioSelleri80}.

Finally, we say that $\Xi$ \emph{violates a Bell inequality} $\{\boldsymbol{\beta},\beta_{\rm L}\}$ if there exists a set $M:=\{M_x\}_{x\in[m]}$ of measurements $M_x\coloneqq\big\{M^{(a)}_x\big\}_{a\in[o]}$, with non-negative measurement operators $M^{(a)}_x$ on $\mathcal{H}$ fulfilling $\sum_{a\in[o]} M^{(a)}_x=\mathbb1$ for all $a\in[m]$, $\mathbb1$ being the identity operator on $\mathcal{H}$, such that the distribution  $\boldsymbol{P}_{\Xi}:=\{P_{\Xi}(a,\boldsymbol{b}|x,\boldsymbol{y})\}
_{a\in[o],\, \boldsymbol{b}\in[\boldsymbol{o}],\, x\in[m],\, \boldsymbol{y}\in[\boldsymbol{m}]}$, defined by
\begin{align}
\label{eq:PofSigma}
P_{\Xi}(a,\boldsymbol{b}|x,\boldsymbol{y})\coloneqq\Tr\left[M^{(a)}_x\,\sigma_{\boldsymbol{b}|\boldsymbol{y}}\right]\ \forall\, \substack{a\in[o],\, \boldsymbol{b}\in[\boldsymbol{o}]\\ x\in[m],\, \boldsymbol{y}\in[\boldsymbol{m}]} ,
\end{align}
violates $\{\boldsymbol{\beta},\beta_{\rm L}\}$, i.e. if $\boldsymbol{\beta}\cdot\boldsymbol{P}_{\Xi}>\beta_{\rm L}$.

Our definition of Bell violation by an assemblage considers only non-sequential measurements on a single copy of the assemblage. For quantum states, it is known that measurements on multiple copies of the state \cite{several,severalIsotropic,networks}, or sequential measurements (filterings) \cite{sequential,sequentialforPOVMs} on a single copy, can produce Bell violations by entangled states that would otherwise yield local correlations. In fact, it has even been suggested \cite{Masanes+08} that every entangled state might be Bell nonlocal in this broader sense. Something similar is expected to happen with steerable assemblages. However, the conditions for Bell violations of assemblages in more general measurement scenarios are outside the scope of the present contribution.

{\it Conditions for Bell violations}.--- Before our first theorem, we need to introduce some notation. Since $\dim(\mathcal H)=2$, any Hermitian operator $O$ on $\mathcal H$ can be decomposed in a Bloch-sphere-like form $O=\frac12\left[\alpha\,\mathbb1+{\boldsymbol r}(O)\cdot{\boldsymbol\sigma}\right]$, where $\alpha\coloneqq\Tr[O]\in\mathbb{R}$, $\boldsymbol\sigma\coloneqq(X,Y,Z)$ is the Pauli-operator vector, and 
\begin{equation}
\label{Bldef}
{\boldsymbol r}(O)\coloneqq\left( \Tr[O\,X],\Tr[O\,Y],\Tr[O\,Z]\right)\in\mathbb{R}^3.
\end{equation}
If $O$ is a state, ${\boldsymbol r}(O)$ represents its Bloch vector.
Finally, for any ${\boldsymbol r}=(x,y,z)\in\mathbb{R}^3$, we denote its Euclidean norm by $\|{\boldsymbol r}\|:=\sqrt{{x}^2+{y}^2+{z}^2}$. 

We can now present our main result, which we prove in the Appendix \ref{app:main}:
\begin{crit}[Criterion for multipartite Bell violations]
\label{crit:conds_for_Bell_viol}
Let $\Xi$ be a generic assemblage given by Eq. \eqref{eq:ensembledef} and let $\{\boldsymbol{\beta},\beta_{\rm L}\}$ be a well-behaved Bell inequality with dichotomic measurements for Alice. Then, $\Xi$ violates $\{\boldsymbol{\beta},\beta_{\rm L}\}$ if, and only if,
\begin{align}
\sum_{x\in[m]}\left[\sum_{ \substack{a\in[2],\, \boldsymbol{b}\in[\boldsymbol{o}]\\ \boldsymbol{y}\in[\boldsymbol{m}]} }
 \frac12\beta_{a,\boldsymbol{b},x,\boldsymbol{y}} \, P(\boldsymbol{b}|\boldsymbol{y}) +\left\|{\boldsymbol s}^\mathrm{opt}_x\right\|\right] > \beta_L, 
\label{eq:main_result}
\end{align}
where
\begin{align}
{\boldsymbol s}^\mathrm{opt}_x\coloneqq{\boldsymbol r}\left(\sum_{\boldsymbol{b}\in[\boldsymbol{o}], \, \boldsymbol{y}\in\boldsymbol{[m]}}
\frac12\left(\beta_{0,\boldsymbol{b},x,\boldsymbol{y}}-\beta_{1,\boldsymbol{b},x,\boldsymbol{y}}\right)\sigma_{\boldsymbol{b}|\boldsymbol{y}}\right)
\label{eq:main_result_r}
\end{align}
Furthermore, the maximal violation is given by von-Neumann measurements along the Bloch-sphere directions ${\boldsymbol s}^\mathrm{opt}_x$, i.e., with 
\begin{equation}
\label{eq:opt_meas}
M^{(a)}_x=\frac{1}{2}\left(\mathbb1 + (-1)^a\, \frac{{\boldsymbol s}^\mathrm{opt}_x}{\|{\boldsymbol s}^\mathrm{opt}_x\|}\cdot\boldsymbol\sigma\right).
\end{equation}  
\end{crit}
Eqs. \eqref{eq:main_result}, \eqref{eq:main_result_r}, and \eqref{eq:opt_meas} are the solutions to optimizations over general (POVM) dichotomic measurements.
The criterion applies to many of the most-widely used Bell inequalities. In the bipartite case, these include, for instance, the CHSH \cite{CHSH} and chained \cite{Pearle70,BraunsteinCaves90} inequalities, as well as the $I_{3322}$ inequality (together with its variants for more outputs for Bob or more inputs for both) \cite{I3322}. In the multipartite scenario, in turn, Criterion \ref{crit:conds_for_Bell_viol} covers important multipartite \cite{MABK} and genuinely multipartite \cite{Svetlichny,generalSvetlichny,Aolita12} Bell inequalities, as well as $(N-1)$-sided DI genuine $N$-partite entanglement witnesses \cite{Bancal,Pal}. For instance, Svetlichny's inequality is obtained by taking $\beta_{a,b_1,b_2,0,0,0}=\beta_{a,b_1,b_2,1,1,1}=(-1)^{a+b_1+b_2} $ and $\beta_{a,b_1,b_2,x,\boldsymbol{y}}=(-1)^{a+b_1+b_2+1} $ otherwise \cite{Svetlichny}; whereas the Svetlichny-like chained inequality introduced in Ref. \cite{Aolita12} is given by $\beta_{a,\boldsymbol{b},x,\boldsymbol{y}}=(-1)^{a+b_1+b_2 + \lfloor (y_2 + x)/m \rfloor + 1}\,\delta_{y_1,[y_2+x]_2}$, where $[a]_2$ stands for $a$ modulo $2$. Violation of the former implies genuinely multipartite nonlocality (GMN), while violation of the latter implies a strong form of GMN that, for large $m$, is a resource for DI quantum secret-sharing protocols against generic nonsignaling  (even post-quantum) eavesdroppers \cite{Aolita12}. Quantum secret sharing (QSS) is an intrinsically multipartite cryptographic protocol with remarkable security properties \cite{QSecretSharing}. Interestingly, unconditional security of QSS has been recently proven in the $(N-1)$-sided DI scenario we consider but in the continuous-variable regime  \cite{Kogias16}.

{\it Conditions for 2-input 2-output bipartite Bell nonlocality}.--- As a crucial application of Criterion \ref{crit:conds_for_Bell_viol}, we focus on the case of two parties, each one with dichotomic inputs and outputs. In this scenario, Bell nonlocality is equivalent to a CHSH violation \cite{Fine}, thus, applying Criterion \ref{crit:conds_for_Bell_viol} to the CHSH inequality, we automatically get a necessary and sufficient condition for nonlocality. This is formalized by the following corollary, whose proof we leave for the Appendix \ref{app:2input2output}. For ease of notation, from now on we omit the subindex 1 from the untrusted party's input $b_1$ and output $y_1$.
\begin{crit}[Criterion for 2-input 2-output bipartite nonlocality]
\label{crit:twotwo}
Let $\Xi$ be a generic assemblage with $m=2$ inputs and $o=2$ outputs per party and $N=2$. Then, $\Xi$ is Bell local if, and only if,
\begin{align}
\label{eq:222}
 \sum_{x\in[2]} \left\|{\boldsymbol t}^\mathrm{opt}_x\right\| \leq 2 \ ,
\end{align}
where
\begin{align}
{\boldsymbol t}^\mathrm{opt}_x\coloneqq{\boldsymbol r}\left(\sum_{b,y\in[2]}(-1)^{b+x\,y}\, \sigma_{b|y}\right). 
\label{eq:222_r}
\end{align}
Furthermore, if inequality \eqref{eq:222} is violated, the maximal violation is given by von Neumann measurements along the Bloch-sphere directions ${\boldsymbol t}^\mathrm{opt}_x$.
\end{crit} 

{\it Connection to steering inequalities}.--- In Ref. \cite{ECavalcanti14} a bipartite steering inequality for correlators was derived. There, it was shown that, if $A_0$ and $A_1$ are any two out of the three Pauli operators in an arbitrary basis of $\mathcal H$ and $\Xi$ is unsteerable, then 
\begin{align}
\label{eq:Australians}
\sqrt{\sum_{x\in[2]}\langle A_x\,(B_0 + B_1)\rangle^2} + 
\sqrt{\sum_{x\in[2]}\langle A_x\,(B_0 - B_1)\rangle^2} \leq 2 \ .
\end{align}
$B_0$ and $B_1$ are unknown, $\pm1$-valued observables of the untrusted part; and $\langle A_x\,B_y\rangle\coloneqq\sum_{a\in[2]}(-1)^a\Tr[\sigma_{b|y}\,A_x]$ for all $x,y\in[2]$. Later on, in Ref. \cite{ECavalcanti16}, the authors found out that a violation of Eq. \eqref{eq:Australians} implies not only that $\Xi$ is steerable but also that it violates the CHSH inequality (under some measurements for Alice not necessarily corresponding to $A_0$ and $A_1$).

\begin{figure}[t]%
\includegraphics[width=0.8\columnwidth]{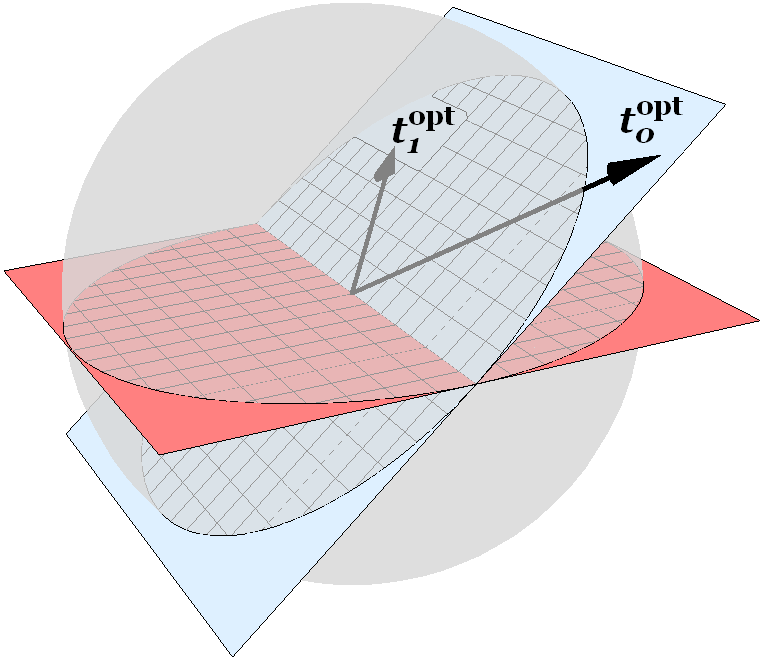}%
\caption{(Color online) Comparison with the steering inequalities of Refs. \cite{ECavalcanti14,ECavalcanti16}. The lhs of inequality \eqref{eq:Australians} is equivalent to that of inequality \eqref{eq:222} evaluated at the projections of ${\boldsymbol t}^\mathrm{opt}_0$ and ${\boldsymbol t}^\mathrm{opt}_1$ onto some fixed chosen plane (represented in red) instead of at ${\boldsymbol t}^\mathrm{opt}_0$ and ${\boldsymbol t}^\mathrm{opt}_1$ themselves. For assemblages for which the plane (represented in light blue) of ${\boldsymbol t}^\mathrm{opt}_0$ and ${\boldsymbol t}^\mathrm{opt}_1$  happens to coincide with the chosen one, \eqref{eq:Australians} are \eqref{eq:222} equivalent. For any other assemblage, inequality \eqref{eq:222} is more effective than inequality \eqref{eq:Australians}.
}
\label{compAust}%
\end{figure}

We can explain this implication in light of Criterion \ref{crit:twotwo}. To this end, we note (see Appendix \ref{app:steerineq}) that Eq. \eqref{eq:222} can be recast as
\begin{align}
\sqrt{\sum_{x\in[3]}\langle A_x\,(B_0 + B_1)\rangle^2} + 
\sqrt{\sum_{x\in[3]}\langle A_x\,(B_0 - B_1)\rangle^2} \leq 2 \ ,
\label{eq:222meas}
\end{align}
where $A_2$ is the third Pauli operator complementary to $A_0$ and $A_1$. Clearly, the left-hand side (lhs) of Eq. \eqref{eq:222meas} is greater or equal than that of Eq. \eqref{eq:Australians}. As a consequence, a violation of Eq. \eqref{eq:Australians} implies a violation \eqref{eq:222meas} and, therefore, of \eqref{eq:222}, consistent with the findings of Ref. \cite{ECavalcanti16}.

The difference, of course, is that a violation of Eq. \eqref{eq:222meas} does not in general imply a violation of Eq. \eqref{eq:Australians}. So, while the former gives a necessary and sufficient condition for 2-input 2-output bipartite Bell nonlocality, the latter only provides a sufficient one.
Interestingly, one can show (see Appendix \ref{app:steerineq}) that, in the basis of $\mathcal H$ in which the Bloch-sphere directions associated to $A_0$ and $A_1$ are contained in the plane of ${\boldsymbol t}^\mathrm{opt}_0$ and ${\boldsymbol t}^\mathrm{opt}_1$, the correlators involving $A_2$ in Eq. \eqref{eq:222meas} vanish and  the lhss of Eq. \eqref{eq:222meas} and Eq. \eqref{eq:Australians} thus coincide (see Fig. \ref{compAust}).
This implies that for every 2-input 2-output bipartite Bell
nonlocal assemblage there exists a pair of mutually unbiased
bases for which the steering in the assemblage is witnessed
via a violation of \eqref{eq:Australians}. The implication was also recently formally proven in Refs. \cite{CostaAngelo15,Quan16} by other reasonings. The pair of bases is readily obtained from Eq. \eqref{eq:222_r}.

 
{\it Concluding remarks}.--- 
We have unambiguously characterized all the $N$-partite  assemblages with a single trusted (qubit) system that can violate --- under general dichotomic trusted measurements on the qubit --- Bell inequalities from a very broad family. 
Indeed, most widely used inequalities are within the firepower of our criterion.
In addition, the optimal POVMs for maximal violation
were also provided and turn out to always be rank-1 projective von Neumman measurements.
Furthermore, for the important particular case of two users with 2 inputs and 2 outputs per user, 
our criterion unambiguously characterizes all Bell nonlocal correlations in the assemblage.
In that setting, we showed that the form of the criterion generalizes that of recently obtained steering inequalities, and sheds light back onto the problem of steering detection within the framework of Bell nonlocality.

Our results hold for the usual scenario of non-sequential measurements on a single copy of the assemblage. Hence, our criterion is to qubit assemblages what the famous analytical criterion of Ref. \cite{Horodecki95} for violation of the CHSH inequality \cite{CHSH} is to 2-qubit states, with the difference that we handle more generic Bell inequalities and in the multipartite case. We leave as an open question the conditions for Bell violations and Bell nonlocality in more general measurement scenarios \cite{sequential,sequentialforPOVMs,several,severalIsotropic,networks,Masanes+08}. 

It is important to remark that the optimizations we have solved can also be solved with semi-definite programming (SDP) \cite{CavalcantiSkrzypczyk16}. However, while SDP can, in general, only give, for each problem instance,  a numeric solution, we provided closed-form analytic expressions for the general case. Analytic solutions both carry more information and are more practical than numeric ones. This is specially relevant in, e.g., security proofs, which are naturally formulated symbolically and rarely admit numeric manipulations. In turn, here, it was precisely having closed-form solutions what made the conceptual connections with steering inequalities possible.

We have developed a practical toolbox to study of the interplay between EPR steering and Bell nonlocality, with potential implications in multipartite cryptographic protocols such as quantum secret sharing \cite{QSecretSharing,Aolita12,Kogias16}. A further interesting prospect would be to explore possible connections with random-number generation schemes that are one-sided DI in that they treat photon sources as black boxes  \cite{Cao2016}. 

{\it Acknowledgements}.--- We thank S. P. Walborn for his support. We acknowledge financial support from the Brazilian agencies CNPq (National Council for Scientific and Technological Development), FAPERJ (Research Support Foundation of the State of Rio de Janeiro), CAPES (Coordination for the Improvement of Higher Education Personnel), and INCT-IQ (National Institute for Science and Technology of Quantum Information).


\appendix

\begin{center}{\large \textbf{Appendix}}\end{center}

\section{Proof of Criterion \ref{crit:conds_for_Bell_viol}}
\label{app:main}
To prove our main theorem, we need to maximize the Bell expression
\begin{equation}
\boldsymbol{\beta}\cdot\boldsymbol{P}_{\Xi} = \sum_{a\in[2],\,\boldsymbol{b}\in[\boldsymbol{o}],\,x\in[m],\,\boldsymbol{y}\in[\boldsymbol{m}]} \beta_{a,\boldsymbol{b},x,\boldsymbol{y}} \, \Tr\left[M^{(a)}_x\,\sigma_{\boldsymbol{b}|\boldsymbol{y}}\right] 
\label{eq:Bell_IneqExclTriv}
\end{equation}
over all sets $M\coloneqq\{M_x\}_{x\in[m]}$ of generic dichotomic POVM measurements $M_x\coloneqq\big\{M^{(0)}_x, M^{(1)}_x\big\}$ by Alice. However, it is known that, to test for Bell nonlocality of quantum states, it suffices to examine only von Neumann (rank-1 projective) measurements \cite{Barrett01}. The same happens for Bell violations by assemblages, which we formalize with the following lemma.
\begin{lem}[Von Neumman measurements are optimal for dichotomic Bell violations]
\label{lemma:excludingPOVMs}
Let $\{\boldsymbol{\beta},\beta_{\rm L}\}$ be an arbitrary well-behaved Bell inequality with dichotomic outputs for Alice and $\Xi$ a generic qubit assemblage. Then, the maximal violation of $\{\boldsymbol{\beta},\beta_{\rm L}\}$ by $\Xi$ is attained under rank-1 projective measurements.
\end{lem}

{\it Proof}.--- Our proof strategy consists in showing that, for an arbitrary assemblage $\Xi$ and generic dichotomic POVM measurements $M$, the distribution $\boldsymbol{P}_{\Xi}$, given by Eq. \eqref{eq:Bell_IneqExclTriv}, obtained from $\Xi$ under $M$, is equivalent to a distribution $\tilde{\boldsymbol{P}}_{\Xi}$ -- obtained from $\Xi$ under a set $\tilde{M}$ of von Neumann measurements -- followed by a local mixing of Alice's outputs. Since $\{\boldsymbol{\beta},\beta_{\rm L}\}$ is well behaved, this implies that if $\boldsymbol{P}_{\Xi}$ violates $\{\boldsymbol{\beta},\beta_{\rm L}\}$, i.e. if $\boldsymbol{\beta}\cdot\boldsymbol{P}_{\Xi}>\beta_{\rm L}$, then $\boldsymbol{\beta}\cdot\tilde{\boldsymbol{P}}_{\Xi}\geq\boldsymbol{\beta}\cdot\boldsymbol{P}_{\Xi}$, which implies that the maximal violation is always attained under von Neumann measurements.

Since, for all $x\in[m]$, the POVM measurement operators $M^{(0)}_x$ and $M^{(1)}_x$ are both non-negative, they can be diagonalized as
\begin{equation}
	\begin{array}{rclcl}
	M_x^{(0)}={} &  \lambda_x^{(0)}  & \Pi_x^{(0)} +{} &   \lambda_x^{(1)}   & \Pi_x^{(1)} \ ,\\
	M_x^{(1)}={} &(1-\lambda_x^{(0)})& \Pi_x^{(0)} +{} & (1-\lambda_x^{(1)}) & \Pi_x^{(1)} \ ,
	\end{array}
\label{eq:Mdecomp}
\end{equation}
where $\Pi_x^{(0)}$ and $\Pi_x^{(1)}$ are rank-1 orthonormal projectors -- i.e. $\Pi_x^{(i)}\Pi_x^{(j)}=\delta_{i,j}\,\Pi_x^{(i)}$, being $\delta_{i,j}$ the Kronecker delta 
 --  acting on $\mathcal{H}$, and where $0\leq\lambda_x^{(a')}\leq1$ for $a'=0,1$. Hence, $\tilde{M}\coloneqq\big\{\Pi_x^{(0)},\Pi_x^{(1)}\big\}_{x\in[m]}$ defines a set of von Neumann measurements.

Substituting Eq.~\eqref{eq:Mdecomp} into Eq.~(\ref{eq:PofSigma}), we find that
\begin{align}
\label{eq:PpovmDecomp}
\nonumber
 &P_{\Xi}(a,\boldsymbol{b}|x,\boldsymbol{y}) = \\  
  &\Tr\left[\left( q(a|0,x) \ \Pi_x^{(0)} +  q(a|1,x) \ \Pi_x^{(1)} \right) \sigma_{\boldsymbol{b}|\boldsymbol{y}}\right] \ ,
\end{align}
where we have introduced
\begin{equation}
q(a|a',x) := \left\{\begin{alignedat}{1} 
                      \lambda_x^{(a')}   \ , & \ \mbox{ if  }\  a=0\\
						 	      1 - \lambda_x^{(a')} \ , & \ \mbox{ if  }\  a=1 \ . 
										\end{alignedat} \right.		
\label{eq:defq}
\end{equation}
Now, defining the distribution $\tilde{\boldsymbol{P}}_{\Xi}$ such that $\tilde{P}_{\Xi}(a',\boldsymbol{b}|x,\boldsymbol{y})\coloneqq\Tr\left[\Pi_x^{(a')}\sigma_{\boldsymbol{b}|\boldsymbol{y}}\right]$, we can write
\begin{equation}
P_{\Xi}(a,\boldsymbol{b}|x,\boldsymbol{y})= \sum_{a'\in[2]} q(a|a',x) \ \tilde{P}_{\Xi}(a',\boldsymbol{b}|x,\boldsymbol{y}) \ .
\label{eq:POVMasLocMix}
\end{equation}
This, as evident from Eq. (\ref{eq:defLocMix}), is the expression of a local mixing of Alice's outputs applied to the von Neumann measurement distribution $\tilde{P}_{\Xi}$. \qed

We can now continue with the proof of Criterion~\ref{crit:conds_for_Bell_viol}. Due to Lemma \ref{lemma:excludingPOVMs}, we need to maximize the Bell expression \eqref{eq:Bell_IneqExclTriv} only over the set of von Neumann measurements, i.e., with measurement operators $M^{(a)}_x$ of the form
\begin{equation}
M^{(a)}_x = \frac{1}{2}\left( \mathbb1 + (-1)^a\,\hat{\boldsymbol{s}}_x\cdot\boldsymbol\sigma\right) \ .
\label{eq:Alice_Projective}
\end{equation}
where each (unit) vector $\hat{\boldsymbol{s}}_x\in\mathbb{R}^3$ represents a direction on the Bloch sphere. Recall that $\boldsymbol\sigma$ is the Pauli-operator vector with respect to a fixed basis of $\mathcal{H}$ of one's preference, so that the vectors $\hat{\boldsymbol{s}}_x$ are the only variables of the optimization.

Using Eqs. \eqref{eq:Bell_IneqExclTriv} and \eqref{eq:Alice_Projective}, and the fact that the vectors $\{\hat{\boldsymbol{s}}_x\}_{x\in[m]}$ are all independent, we get
\begin{widetext}
\begin{equation}
\max_{\{\hat{\boldsymbol{s}}_x\}}\boldsymbol{\beta}\cdot\boldsymbol{P}_{\Xi}=
\sum_{x\in[m]} \left\{\sum_{a\in[2],\,\boldsymbol{b}\in[\boldsymbol{o}],\,\boldsymbol{y}\in[\boldsymbol{m}]}\frac12\beta_{a,\boldsymbol{b},x,\boldsymbol{y}} \, P(\boldsymbol{b}|\boldsymbol{y}) +\max_{\{\hat{\boldsymbol{s}}_x\}}\Tr\left[
B_x\, \hat{\boldsymbol{s}}_x \cdot \boldsymbol{\sigma}\right]\right\}\ .
\label{eq:RecastExpr}
\end{equation}
\end{widetext}
where, for each $x\in[m]$, we have introduced the Hermitean operator on $\mathcal{H}$
\begin{equation}
B_x \coloneqq \sum_{\boldsymbol{b}\in[\boldsymbol{o}],\,\boldsymbol{y}\in[\boldsymbol{m}]} \frac12(\beta_{0,\boldsymbol{b},x,\boldsymbol{y}}-\beta_{1,\boldsymbol{b},x,\boldsymbol{y}})\, \sigma_{\boldsymbol{b}|\boldsymbol{y}} \ .
\label{eq:BobOps}
\end{equation}
Note that $B_x$ coincides with the expression inside the brackets of Eq. (\ref{eq:main_result_r}).
Then, using that $\Tr[B_x  \, \hat{\boldsymbol{s}}_x\cdot\boldsymbol{\sigma}] = \boldsymbol r(B_x)\cdot\hat{\boldsymbol{s}}_x$, with the vector function $\boldsymbol r$ defined in Eq. (\ref{Bldef}), the maximization is finally reduced to
\begin{equation}
\max_{\{\hat{\boldsymbol{s}}_x\}}\Tr\left[
B_x\, \hat{\boldsymbol{s}}_x \cdot \boldsymbol{\sigma}\right]=
\max_{\{\hat{\boldsymbol{s}}_x\}}\ \boldsymbol{r}(B_x)\cdot {\hat{\boldsymbol{s}}}_x \ .
\label{eq:ReducedProb}
\end{equation}
Clearly, the maximum is
\begin{equation}
\max_{\{\hat{\boldsymbol{s}}_x\}}\Tr\left[
B_x\, \hat{\boldsymbol{s}}_x \cdot \boldsymbol{\sigma}\right]=\|{\boldsymbol s}^{\rm opt}_x\| \ ,
\label{BestExpr}
\end{equation}
with
\begin{equation}
\label{s_Ay}
{\boldsymbol s}^\mathrm{opt}_x=\boldsymbol{r}(B_x)
\end{equation}
 for all $x\in[m]$, attained by
\begin{equation}
{\hat{\boldsymbol{s}}}_x = \frac{ {\boldsymbol s}_x^{\rm opt} }{\| {\boldsymbol s}_x^{\rm opt} \|} \ .
\label{BestChoice}
\end{equation}
Substituting Eq. \eqref{BestChoice} into Eq. \eqref{eq:Alice_Projective}, one obtains the optimal measurement settings of Eq.~(\ref{eq:opt_meas}). Using, in turn, Eqs. \eqref{BestExpr} and \eqref{s_Ay}, one sees that Eq.~\eqref{eq:RecastExpr} is equivalent to the left-hand side of Eq.(\ref{eq:main_result}).

\section{Proof of Criterion \ref{crit:twotwo}}
\label{app:2input2output}
In the 2-input, 2-output scenario, by virtue of Fine's theorem \cite{Fine}, Bell nonlocality is equivalent to the violation of the CHSH inequality, given by
\begin{equation}
\langle A_0\,B_0 \rangle + \langle A_0\,B_1 \rangle + \langle A_1\,B_0 \rangle - \langle A_1\,B_1 \rangle \leq 2 \ ,
\label{eq:CHSHexplicit}
\end{equation}
or any of its 8 symmetries (defined by swapping around the minus sign with the other terms, by applying an overall sign change, or by doing both). So, it suffices to show that a violation of Eq. (\ref{eq:222}) is equivalent to the violation of any of the 8 symmetries of the CHSH inequality.

In the notation of Eq.~(\ref{eq:Bell_Ineq}), and omitting the subindices from $b_1$ and $y_1$, the CHSH inequality \eqref{eq:CHSHexplicit} is expressed as
\begin{equation}
\beta_{a,b,x,y}=(-1)^{a+b}(-1)^{x\,y} \text{ and }\beta_{\rm L}=2,
\label{eq:CHSHbeta}
\end{equation}
with $m=2=o$. Its symmetries, in turn, are obtained by replacing $x$ or $y$ by their negations modulo 2, by applying an overall sign change to $\{\boldsymbol\beta,\beta_{\rm L}\}$, and by applying any composition of the three.

Substituting Eq. \eqref{eq:CHSHbeta} in Eqs. (\ref{eq:main_result}) and (\ref{eq:main_result_r}) leads to Eqs. (\ref{eq:222}) and (\ref{eq:222_r}), as the reader can straightforwardly verify. This shows that the violation of Eq. (\ref{eq:222}), with the measurement direction $\boldsymbol t_x^{\rm opt}$ given by Eq. (\ref{eq:222_r}), is equivalent to the violation of the CHSH inequality \eqref{eq:CHSHexplicit}. Now, note that any of the other symmetries mentioned above either does not explicitly introduce any change in Eqs. (\ref{eq:222}) and (\ref{eq:222_r}) or simply corresponds to the relabelings $\boldsymbol t_0^{\rm opt}\leftrightarrow\boldsymbol t_1^{\rm opt}$, $\boldsymbol t_x^{\rm opt}\to-\boldsymbol t_x^{\rm opt}$, for all $x\in[2]$, or compositions of the two. None of the latter alters the statements of Criterion \ref{crit:twotwo}. That is, the violation of Eq. (\ref{eq:222}), with $\boldsymbol t_x^{\rm opt}$ given by Eq. (\ref{eq:222_r}), is equivalent to the violation of any of the symmetries of the CHSH inequality, which finishes the proof.

\section{Equivalence between Eqs. (\ref{eq:222}) and (\ref{eq:222meas})}
\label{app:steerineq}
For any assemblage $\Xi\coloneqq\{\sigma_{b|y}\}_{b,y\in[2]}$, the correlator $\langle A_\alpha \,B_y\rangle$, where $B_y$ is a $\pm1$-valued unknown observable of Bob's subsystem and $A_{\alpha}$ is a Pauli operator on $\mathcal H$, is given by
\begin{align}
\nonumber
\langle A_\alpha \, B_y\rangle & = \sum_{b\in[2]} (-1)^{b} \Tr \left( \sigma_{b|y} \, A_{\alpha} \right)\\
& = \boldsymbol r \left(\sum_{b\in[2]} (-1)^{b} \, \sigma_{b|y} \right)\cdot \hat{\boldsymbol v}_{\alpha} \ ,
\label{eq:corrbasic2}
\end{align}
where $\hat{\boldsymbol v}_{\alpha}$ is a unit vector in the Bloch sphere in the direction of $A_\alpha$. Using Eqs. \eqref{eq:corrbasic2}, (\ref{eq:222_r}) and the linearity of the vector function $\boldsymbol r$, one sees that 
\begin{equation}
{\boldsymbol t}^\mathrm{opt}_0\cdot \hat{\boldsymbol v}_{\alpha} =  \langle A_\alpha \, (B_0+B_1)\rangle
\label{eq:corrascomponent0}
\end{equation}
and, analogously,
\begin{equation}
{\boldsymbol t}^\mathrm{opt}_1\cdot \hat{\boldsymbol v}_{\alpha}=  \langle A_\alpha \, (B_0-B_1)\rangle \ .
\label{eq:corrascomponent1}
\end{equation}
It is now straightforward to see, from the definition of the Euclidian norm, that Eq.~(\ref{eq:222}) is equivalent to Eq.~(\ref{eq:222meas}) for $\{A_0,A_1,A_2\}$ the Pauli operators in any orthonormal basis of $\mathcal{H}$. Furthermore, one can also see that the lhs of inequality (\ref{eq:Australians}) is equivalent to the lhs of 
inequality (\ref{eq:222}) evaluated at the projections of ${\boldsymbol t}^\mathrm{opt}_0$ and ${\boldsymbol t}^\mathrm{opt}_1$ onto the plane orthogonal to $\hat{\boldsymbol v}_{2}$, instead of at ${\boldsymbol t}^\mathrm{opt}_0$ and ${\boldsymbol t}^\mathrm{opt}_1$ themselves.



\end{document}